\documentclass[aps,prd,preprint,amsmath,amssymb,nofootinbib]{revtex4}
\usepackage{amsfonts}
\usepackage{mathrsfs}
\usepackage{graphicx}
\usepackage{subfigure}
\usepackage{epsfig}
\usepackage{graphicx}
\usepackage{slashed}
\usepackage{color}
\usepackage{float}
\usepackage[normalem]{ulem}
\newcommand{\bqa}{\begin{eqnarray}}
\newcommand{\eqa}{\end{eqnarray}}
\newcommand{\beq}{\begin{equation}}
\newcommand{\eeq}{\end{equation}}
\allowdisplaybreaks[1]
\graphicspath{{fig/}{dia/}} \DeclareGraphicsExtensions{.eps}
\begin{document}

\title{The S-wave topped meson\\[0.7cm]}

\author{Jun-Hao Zhang$^{1,2}$}
\author{Shuo Yang$^{1,2}$}
\email{shuoyang@lnnu.edu.cn}
\author{Bing-Dong Wan$^{1,2}$}
\email{wanbd@lnnu.edu.cn} 
\vspace{+3pt}

\affiliation{$^1$School of Physics and Electronic Technology, Liaoning Normal University, Dalian 116029, China\\
$^2$ Center for Theoretical and Experimental High Energy Physics, Liaoning Normal University, Dalian 116029, China
}

\begin{abstract}
\vspace{0.5cm}
Motivated by the recent near-threshold enhancement in top-quark pair production reported by CMS and ATLAS, we study the S-wave spectral structure of heavy-light systems containing a single top quark, namely $t\bar{q}$, $t\bar{c}$, and $t\bar{b}$, within the instantaneous Bethe-Salpeter formalism. Because the top quark decays on a timescale much shorter than the typical hadronization time, the discrete eigenvalues we obtain should be interpreted as model-dependent reference positions of possible quasi-bound heavy-light configurations, rather than as predictions for fully formed conventional hadrons. The numerical results indicate that the masses of these configurations lie close to the top-quark mass.  
For the $t\bar{b}$ system, the masses of the first four S-wave $0^{-}$ radial states are about $5.1$, $5.4$, $5.6$, and $5.7$~GeV above the top-quark mass, respectively. For the $t\bar{c}$ system, the corresponding values are about $1.9$, $2.2$, $2.5$, and $2.6$~GeV. We also briefly discuss possible production and decay patterns at a qualitative level, which may serve as a reference for future dedicated phenomenological studies or for experimental constraints.
\end{abstract}
\pacs{11.55.Hx, 12.38.Lg, 12.39.Mk} \maketitle
\newpage

\section{Introduction}\label{sec:1}
Owing to its large mass and decay width 
($1.42^{+0.19}_{-0.15}$ GeV) \cite{1}, the top quark decays on a timescale 
($\tau_t\approx 5 \times 10^{-25}\;s$) much shorter than the characteristic 
QCD hadronization time, $\tau_{\text{had}}\sim 1/\Lambda_{\text{QCD}}\approx 10^{-23}\;s$. 
This hierarchy of timescales was already pointed out in Ref.~\cite{Bigi:1986jk} to imply that the top quark decays before hadronization can occur, 
leading to important phenomenological consequences. 
As a result, the top quark is qualitatively different from the lighter quarks that hadronize on QCD timescales, making it a useful probe of the interplay between perturbative and nonperturbative strong-interaction dynamics.

Top quark pairs are produced near threshold at the LHC, providing an opportunity to explore strong-interaction effects in the presence of very heavy quarks. Recently, both the CMS and ATLAS collaborations have reported a statistically significant excess in the $t\bar{t}$ invariant-mass spectrum near threshold \cite{2,3}, consistent with a pseudoscalar toponium state. Similar structures in earlier ATLAS and CMS datasets also support this interpretation \cite{4,5}. Motivated by these developments, a wide range of theoretical investigations has been carried out~\cite{6,7,8,9,10,11,12,13,14,15,16,17,18,19,20,23,24,25,26,27,28,29,30,31}, including studies of toponium production and decay~\cite{25,26,27,28} and possible ``topped'' mesons and baryons~\cite{29,30,31}.
In addition, numerous earlier theoretical studies had already anticipated the existence of toponium and analyzed its threshold properties, spectrum, and collider signatures~\cite{32,33,35,36,38,39,40,41,42,43,44,45,46,47,48,49,50,51,56,57,59}.

In light of these developments, it is interesting to ask whether analogous heavy-light structures associated with a single top quark can be meaningfully discussed at the level of model spectroscopy. Unlike toponium, where the dominant binding scale is set by the heavy-quark pair near threshold, a topped heavy-light system involves competition between the short top lifetime and a binding scale of hadronic size. This makes the physical observability of such configurations highly uncertain. Nevertheless, an exploratory calculation of the corresponding heavy-light spectral structure may still be useful as a theoretical benchmark, provided its limitations are explicitly acknowledged.

In the present work we adopt the instantaneous Bethe-Salpeter (Salpeter) approximation with constituent-type quark propagators. This framework is widely used in meson spectroscopy and provides a convenient relativistic setup for comparing different flavor combinations. At the same time, it does not incorporate the finite width of the top quark explicitly in the propagator, interaction kernel, or spectral representation, and it does not aim at a symmetry-preserving treatment with fully dressed propagators. Therefore, the discrete eigenvalues obtained below should not be interpreted as predictions of conventional long-lived topped hadrons, but rather as model-dependent reference positions of possible quasi-bound heavy-light configurations.

With this interpretation, the study of topped mesons offers a means to probe the boundary between perturbative and nonperturbative QCD in the presence of an unstable heavy quark. Should future measurements find hints of such top-light correlations, model studies of this type could help systematize the relevant mass scales and decay patterns.

In this work, we investigate the S-wave spectral structure of systems containing a single top quark, namely $t\bar{q}$, $t\bar{c}$, and $t\bar{b}$, using the instantaneous Bethe-Salpeter (BS) formalism. Section~\ref{sec:2} briefly introduces the formalism and the inputs used in our calculation. Section~\ref{sec:3} presents the numerical results. Section~\ref{sec:4} contains a brief qualitative discussion of possible production and decay patterns. Section~\ref{sec:5} summarizes the main points.

\section{Bethe-Salpeter Equation}\label{sec:2}

In quantum field theory, the BS equation provides a relativistic framework for the description of two-body bound states~\cite{Bethe1951,Salpeter1952,Lurie1968,Itzykson1980}. In the present work, we employ its instantaneous Salpeter version together with constituent-type quark propagators; therefore, the calculation should be regarded as a phenomenological model study rather than a fully symmetry-preserving Bethe-Salpeter treatment. The BS wave function of a quark--antiquark bound state is defined by the time-ordered vacuum-to-bound-state matrix element.
\begin{equation}
\label{e1}
\chi(x_1,x_2)\;=\;\langle 0\,|\,T\{\psi(x_1)\,\bar{\psi}(x_2)\}\,|P\rangle,
\end{equation}
where $x_1$ and $x_2$ denote the spacetime coordinates of the quark and antiquark, respectively, $|P\rangle$ is the bound-state eigenvector with total four-momentum $P$, and $T$ denotes the time-ordering operator. Eq.~\eqref{e1} serves as the starting point for deriving the BS equation in momentum space, after performing a Fourier transformation and imposing an appropriate interaction kernel.

The BS wave function in momentum space is defined as the Fourier transform of the coordinate-space wave function,
\begin{equation}
\label{e2}
\chi_P(q)
= e^{-i P \cdot X} \int d^{4}x \, e^{-i q \cdot x} \, \chi(x_1,x_2),
\end{equation}
where $q$ denotes the relative four-momentum between the quark and the antiquark. The ``center-of-mass'' coordinate $X$ and the ``relative'' coordinate $x$ are defined as
\begin{equation}
\label{eq:COM_relative}
X = \frac{m_1}{m_1 + m_2} \, x_1 
  + \frac{m_2}{m_1 + m_2} \, x_2,
\qquad
x = x_1 - x_2,
\end{equation}
With $m_1$ and $m_2$ denoting the masses of the quark and antiquark, respectively, the bound-state BS equation in momentum space takes the form
\begin{equation}
\label{e3}
S^{-1}_1(p_1)\chi_{_P}(q)S^{-1}_2(-p_2)=i\int\frac{d^4k}{(2\pi)^4} V(P; q,k)\chi_{_P}(k).
\end{equation}
Here the fermion propagators are given by
\begin{equation}
S_i(\pm p_i) = \frac{i}{\pm \slashed{p}_i - m_i}, \qquad (i=1,2),
\end{equation}
I notice the provided text appears to be an incomplete fragment — it ends mid-sentence. Could you please supply the full LaTeX section you would like me to proofread? That will allow me to ensure a coherent and accurate revision.
\begin{equation}
\label{e4}
p_i = \frac{m_i}{m_1 + m_2} \, P + Jq,
\end{equation}
where $J = 1$ for the quark ($i=1$) and $J = -1$ for the antiquark ($i=2$). With the definitions
\begin{equation}
p_{i_P} \equiv \frac{P \cdot p_i}{M}, \qquad
p_{i\perp}^{\mu} \equiv p_i^{\mu} - \frac{P \cdot p_i}{M^{2}} P^{\mu},
\end{equation}
The propagator $S_i(J p_i)$ can be decomposed as
\begin{equation}
\label{eq:prop_decompose}
-i J\, S_i(J p_i)
= \frac{\Lambda_i^{+}(q_\perp)}{p_{i_P} - \omega_i + i \epsilon}
+ \frac{\Lambda_i^{-}(q_\perp)}{p_{i_P} + \omega_i - i \epsilon},
\end{equation}
where the projection operators are defined by
\begin{equation}
\label{eq:projector}
\begin{aligned}
\Lambda_i^{\pm}(q_\perp) 
&\equiv \frac{1}{2 \omega_i}
\left[
\frac{\slashed{P}}{M}\, \omega_i
\pm \left(\slashed{p}_{i\perp} + J\, m_i \right)
\right],
\\[4pt]
\omega_i 
&\equiv \sqrt{m_i^2 - p_{i\perp}^{\,2}} \, .
\end{aligned}
\end{equation}

Under the instantaneous approximation, the interaction kernel in the center-of-mass frame can be written as
\begin{equation}
V(P;q,k)\big|_{\vec{P}=0} \simeq V(q_\perp,k_\perp).
\end{equation}
Accordingly, the BS equation can be reduced to
\begin{equation}
\label{eq:BS_reduce}
\chi_{P}(q)= S_1(p_1)\,\eta_{P}(q_\perp)\,S_2(-p_2),
\end{equation}
with
\begin{equation}
\label{eq:eta_def}
\eta_{P}(q_\perp)
= \int \frac{d^3 k_\perp}{(2\pi)^3}\,
V(q_\perp,k_\perp)\,
\varphi_{P}(k_\perp),
\end{equation}
where the three-dimensional BS wave function is defined as
\begin{equation}
\label{eq:Salpeter_wf}
\varphi_{P}(q_\perp)
\equiv i \int \frac{d q_{P}}{2\pi}\,\chi_{P}(q).
\end{equation}
By introducing the projected components of the Salpeter wave function as
\begin{equation}
\label{eq:Salpeter_projected}
\varphi^{\pm\pm}_{P}(q_\perp)
\equiv 
\Lambda_1^{\pm}(q_\perp) \, \frac{\slashed{P}}{M} \, 
\varphi_P(q_\perp) \, \frac{\slashed{P}}{M} \, \Lambda_2^{\pm}(q_\perp),
\end{equation}
The three-dimensional BS wave function can be decomposed into its four components:
\begin{equation}
\varphi_P(q_\perp) 
= \varphi_P^{++}(q_\perp) + \varphi_P^{+-}(q_\perp) 
+ \varphi_P^{-+}(q_\perp) + \varphi_P^{--}(q_\perp).
\end{equation}
Correspondingly, the BS equation \eqref{eq:BS_reduce} can be decomposed into four coupled equations for the projected components:
\begin{align}
\label{eq:varphi_pp}
(M - \omega_1 - \omega_2)\, \varphi_P^{++}(q_\perp) 
&= \Lambda_1^+(q_\perp)\, \eta_P(q_\perp)\, \Lambda_2^+(q_\perp), \\[2mm]
\label{eq:varphi_mm}
(M + \omega_1 + \omega_2)\, \varphi_P^{--}(q_\perp) 
&= - \Lambda_1^-(q_\perp)\, \eta_P(q_\perp)\, \Lambda_2^-(q_\perp), \\[1mm]
\label{eq:varphi_pm_mp}
\varphi_P^{+-}(q_\perp) &= \varphi_P^{-+}(q_\perp) = 0,
\end{align}
where $\omega_i = \sqrt{m_i^2 + q_\perp^2}$ and $\Lambda_i^\pm(q_\perp)$ are the projection operators defined in Eq.~\eqref{eq:projector}.

To solve the BS equation, a thorough understanding of the interquark potential is essential. According to lattice QCD calculations, the static heavy quark--antiquark potential is accurately described by a long-range linear confining potential (Lorentz scalar $V_S$) and a short-range one-gluon-exchange potential (Lorentz vector $V_V$) \cite{61,62,63}:
\begin{align}
\label{eq:potential_r}
V(r) &= V_S(r) + \gamma_\mu \otimes \gamma^\mu \, V_V(r), \nonumber \\
V_S(r) &= \lambda r \, \frac{1 - e^{-\alpha r}}{\alpha r} , \nonumber \\
V_V(r) &= - \frac{4}{3} \frac{\alpha_s(r)}{r} \, e^{-\alpha r}.
\end{align}
Here, the factor $e^{-\alpha r}$ is introduced not only to avoid infrared divergence but also to parametrize, at the model level, possible screening effects in the interquark potential \cite{64,Born1989}. In the present topped-meson context, this factor should be viewed as part of the potential-model dependence, rather than as a literal dynamical realization of additional light degrees of freedom.
\begin{align}
\label{eq:potential_p}
V(\vec{p}) &= (2\pi)^3 V_S(\vec{p}) + \gamma_\mu \otimes \gamma^\mu \, (2\pi)^3 V_V(\vec{p}), \nonumber \\
V_S(\vec{p}) &= - \Big(\frac{\lambda}{\alpha}  \Big) \delta^3(\vec{p}) 
+ \frac{\lambda}{\pi^2} \frac{1}{(\vec{p}^{\,2} + \alpha^2)^2}, \nonumber \\
V_V(\vec{p}) &= - \frac{2 \alpha_s}{3 \pi^2} \frac{1}{\vec{p}^{\,2} + \alpha^2}.
\end{align}
Here, $\alpha_s$ is the strong coupling constant, and the parameters $\alpha$ and $\lambda$ characterize the potential. 
Ref.~\cite{35} indicates that the mass splitting between the S-wave $J^P=0^-$ and $1^-$ states is negligible due to the large top-quark mass, i.e.,
\begin{equation}\label{appro}
M_{0^{-}}(n) = M_{1^{-}}(n).
\end{equation}
For a heavy-light system, the physical classification is more naturally expressed in terms of the heavy-quark spin and the total angular momentum of the light degrees of freedom. In what follows, we therefore present the states primarily using their quantum numbers $J^P=0^-$ and $1^-$, while retaining the older notation only when useful for comparison with the numerical implementation and related literature. We study the S-wave top-meson system up to $n=4$ and calculate both the $0^{-}$ and $1^{-}$ states.

\section{Numerical evaluation}\label{sec:3}

To solve the BS equation numerically, the relevant parameters must be specified.  
These parameters are determined phenomenologically by fitting to experimental data such that the calculated meson masses are consistent with those reported by the Particle Data Group (PDG)~\cite{1}. The top-quark mass is taken as an input parameter, set to the PDG central value, and no dedicated analysis of the mass-scheme dependence is attempted.  
The resulting parameter values are as follows:
\begin{align}
&m_b=4.960\ {\rm GeV},\quad m_c=1.620\ {\rm GeV},\quad m_s=0.500\ {\rm GeV},\quad m_d=0.311\ {\rm GeV},\nonumber\\
&m_u=0.305\ {\rm GeV},\quad \alpha_s(m_t)=0.11,\quad\alpha=0.06\ {\rm GeV},\quad \lambda=0.18\ {\rm GeV}^2. \nonumber
\end{align}
Based on the formalism and parameters outlined above, we compute the masses of the $t\bar{q}$, $t\bar{c}$, and $t\bar{b}$ states up to the fourth radial excitation ($n=4$). The corresponding numerical results are shown in Table~\ref{tab1} for the S-wave $J^P=0^-$ states.

\begin{table}[htbp]
  \centering
  \caption{Masses (in units of 100~GeV) of S-wave $J^P=0^-$ states for various quark combinations.}
  \label{tab1}  
\begin{tabular}{|c|c|c|c|c|c|}
    \toprule
    State & $t\bar{b}$ & $t\bar{c}$ & $t\bar{s}$ & $t\bar{d}$ & $t\bar{u}$ \\\hline
    ${1S\,(0^-)}$ & $1.778 $ & $1.747 $ & $1.736 $ & $1.734 $ & $1.734 $ \\ \hline
    ${2S\,(0^-)}$ & $1.781 $ & $1.750 $ & $1.740 $ & $1.738 $ & $1.738 $ \\\hline
    ${3S\,(0^-)}$ & $1.783 $ & $1.752 $ & $1.742 $ & $1.739 $ & $1.739 $ \\\hline
    ${4S\,(0^-)}$ & $1.785 $ & $1.754 $ & $1.743 $ & $1.741 $ & $1.741 $ \\\hline
 \hline
  \end{tabular}
\end{table}

From our calculations of the S-wave $J^P=1^-$ states, we find that, for identical quark configurations and principal quantum number $n$, the masses of the $0^-$ and $1^-$ states coincide within our numerical accuracy, confirming that Eq.~(\ref{appro}) is a good approximation in the present model.

\begin{figure}
\includegraphics[width=6.8cm]{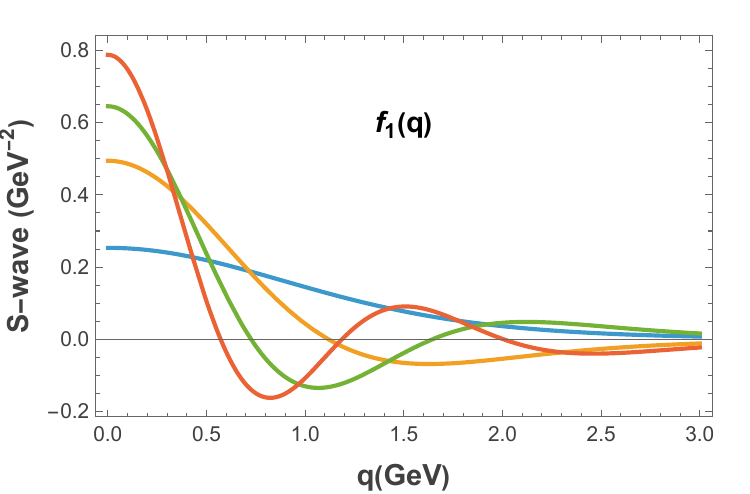}
\includegraphics[width=6.8cm]{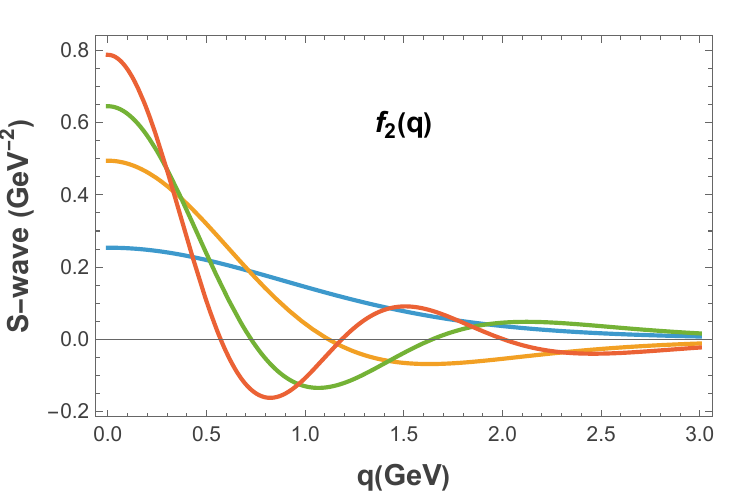}
\caption{Radial components $f_1(q)$ and $f_2(q)$ of the Salpeter wave function for the S-wave $J^P=0^-$ $t\bar{b}$ system: the blue, orange, green, and red curves correspond to the first four radial excitations.} \label{figtb0-}
\end{figure}

As an example, we present explicitly the BS wave function of the $t\bar{b}$ $1S\,(0^-)$ state; the wave functions for the other states are collected in Appendix~\ref{BSwave0-}. Following Refs.~\cite{65,66}, for $0^{-}$ states the Salpeter wave function can be decomposed into two radial components, $f_1(q)$ and $f_2(q)$, which are parts of the full Salpeter amplitude $\varphi_P(q_\perp)$. The explicit decomposition formulas are given in Refs.~\cite{65,66}. We show the corresponding curves for the $t\bar{b}$ system in Fig.~\ref{figtb0-}, mainly to illustrate the qualitative nodal structure of the numerical solutions.

\section{Possible Production and Decay Modes of Topped Mesons}\label{sec:4}

Within the Standard Model, the lifetime of the top quark is extremely short,
\begin{equation}
\tau_t \simeq \frac{1}{\Gamma_t} \sim 5 \times 10^{-25}~\text{s},
\end{equation}
which is significantly shorter than the typical hadronization timescale,
\begin{equation}
\tau_{\text{had}} \sim 10^{-23}~\text{s}.
\end{equation}
Thus, a free top quark decays weakly before it can hadronize into a conventional color-singlet bound state. For this reason, the topped-meson configurations studied in the present work should be interpreted cautiously: within our instantaneous Bethe-Salpeter framework, they represent model-dependent, quasi-bound heavy-light configurations, not ordinary hadrons with fully developed nonperturbative dressing.

In this section, we present only a brief qualitative discussion of the possible production mechanisms and decay modes of such configurations. A quantitative treatment of production rates, line shapes, branching fractions, and backgrounds is beyond the scope of the present work.

At hadron colliders such as the LHC, the dominant production mechanisms for top quarks are
\begin{equation}
pp \to t\bar{t} + X, \qquad pp \to t(\bar{t}) + X,
\end{equation}
Via gluon-gluon fusion and quark-antiquark annihilation.

A topped configuration $T \equiv t\bar{q}$ (or $t\bar{c}$, $t\bar{b}$) may be viewed schematically as arising when a produced top quark becomes correlated with a nearby antiquark before decaying. Representative partonic subprocesses include:
\begin{align}
gg &\to t\bar{t}, \quad t + \bar{q} \longrightarrow (t\bar{q}), \\
q\bar{q} &\to t\bar{t}, \quad t + \bar{q} \longrightarrow (t\bar{q}), \\
gb &\to tW^{-}, \quad t + \bar{b} \longrightarrow (t\bar{b}).
\end{align}

Among these possibilities, the $t\bar{b}$ system may have a larger overlap than the lighter channels because of the heavier $\bar{b}$ quark. We stress, however, that no quantitative production estimate is attempted here.

Schematically, the production amplitude may be written as
\begin{equation}
\mathcal{M}(pp \to T + X) \sim \mathcal{M}(pp \to t + \bar{q} + X)\, \psi_{T}(0),
\end{equation}
where $\psi_{T}(0)$ denotes a symbolic overlap factor characterizing the compactness of the configuration. It is introduced here only as a qualitative indicator, not as an explicitly calculated Schr\"odinger wave function at the origin.

Illustrative production mechanisms include:
\begin{itemize}
  \item Gluon-gluon fusion: two gluons produce a $t\bar{t}$ pair, and the $t$ quark subsequently binds with a light antiquark from the QCD vacuum.
  \item Quark-antiquark annihilation: $q\bar{q} \to t\bar{t}$, followed by $t\bar{q}$ binding to form a $(t\bar{q})$ bound state.
  \item Associated production: $gb \to tW^{-}$, where the final $t$ and $\bar{b}$ form a $(t\bar{b})$ state.
\end{itemize}

These processes are strongly suppressed by the extremely short lifetime of the top quark; consequently, the production cross section of topped configurations is expected to be very small. We do not attempt a quantitative estimate in the present work. If formed, such a topped configuration is primarily unstable due to the weak decay of the top quark inside the system:
\begin{equation}
t \to W^{+} b.
\end{equation}
As a result, the dominant decay channel of a topped meson can be written as
\begin{equation}
(t\bar{q}) \to W^{+} + b + \bar{q}.
\end{equation}
Depending on the decay mode of the $W^{+}$ boson, one further obtains
\begin{align}
(t\bar{q}) &\to \ell^{+} \nu_{\ell} + b + \bar{q}, \\
(t\bar{q}) &\to q\bar{q}' + b + \bar{q}.
\end{align}

For the $t\bar{b}$ system, an alternative decay channel is
\begin{equation}
(t\bar{b}) \to W^{+} + b + \bar{b},
\end{equation}
followed by $b\bar{b}$ hadronization into bottom-flavored final states.

In addition, highly excited states may undergo cascade transitions through
electromagnetic or strong processes before the weak decay of the top quark, such as
\begin{align}
(t\bar{q})^{*} &\to (t\bar{q}) + \gamma, \\
(t\bar{q})^{*} &\to (t\bar{q}) + g,
\end{align}
Although such transitions are expected to be strongly suppressed due to the short lifetime of the top quark, at the qualitative level one may search for top-light correlations in final states containing
\begin{itemize}
\item a high-$p_T$ $W$ boson,
\item one or two $b$-jets,
\item a possible additional light jet from the spectator antiquark.
\end{itemize}

Accordingly, a possible signal topology may be summarized as
\begin{equation}
pp \to T + X \to W^{+} + b + \bar{q} + X,
\end{equation}
This might affect invariant-mass distributions or angular correlations relative to standard top backgrounds. Determining whether such effects are observable requires a dedicated quantitative analysis.

At present, this discussion should be regarded only as a qualitative phenomenological guide. More realistic predictions would require including the finite top width along with explicit signal and background estimates.

\section{Summary}\label{sec:5}
In this work, we investigate the S-wave spectral structure of systems containing a single top quark, including the $t\bar{q}$, $t\bar{c}$, and $t\bar{b}$ channels, within the framework of the Bethe-Salpeter equation under the instantaneous approximation. We obtain the mass spectra up to the $n=4$ states; the numerical results are presented in Tab.~\ref{tab1}.

Since the top quark decays on a timescale much shorter than the typical hadronization time, the resulting discrete eigenvalues should not be interpreted as predictions of ordinary long-lived hadrons containing a top quark. Instead, within the present model, they serve as reference positions for possible quasi-bound heavy-light configurations. We also provide a brief qualitative discussion of possible production and decay patterns, but emphasize that quantitative phenomenology, including finite-width effects and realistic background studies, lies beyond the scope of this work. In this sense, the present study offers a model benchmark for future, more complete investigations of top-flavored heavy-light systems.

\vspace{.5cm} {\bf Acknowledgments} \vspace{.5cm}

This work was supported in part by the National Natural Science Foundation of China under Grants 12575106 and 12147214, and the Specific Fund of Fundamental Scientific Research Operating Expenses for Undergraduate Universities in Liaoning Province under Grant No. LJ212410165019.

\bibliographystyle{unsrt}
\bibliography{ref}

\begin{widetext}
\appendix
\section{The BS wave functions for S-wave $J^P=0^-$ states}\label{BSwave0-}
The BS wave functions for the S-wave $J^P=0^-$ $t\bar{c}$, $t\bar{s}$, $t\bar{d}$, and $t\bar{u}$ systems are shown in Figs.~\ref{figtc0-}-\ref{figtu0-}. These figures are included mainly to document the qualitative structure of the numerical solutions.

\begin{figure}[H]
\includegraphics[width=6.8cm]{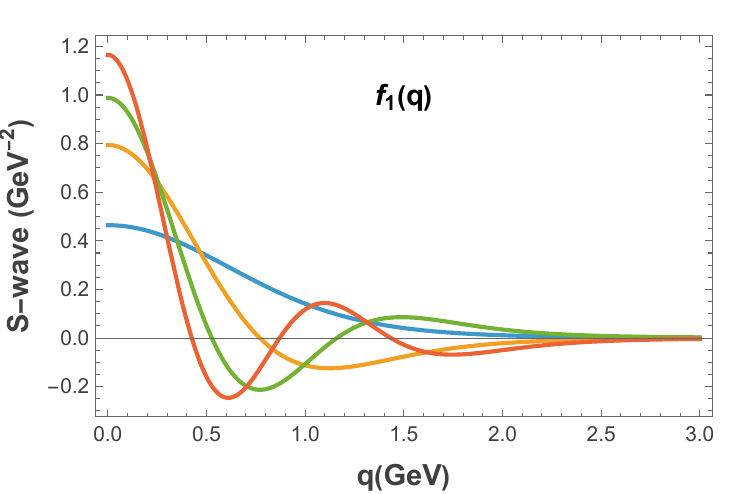}
\includegraphics[width=6.8cm]{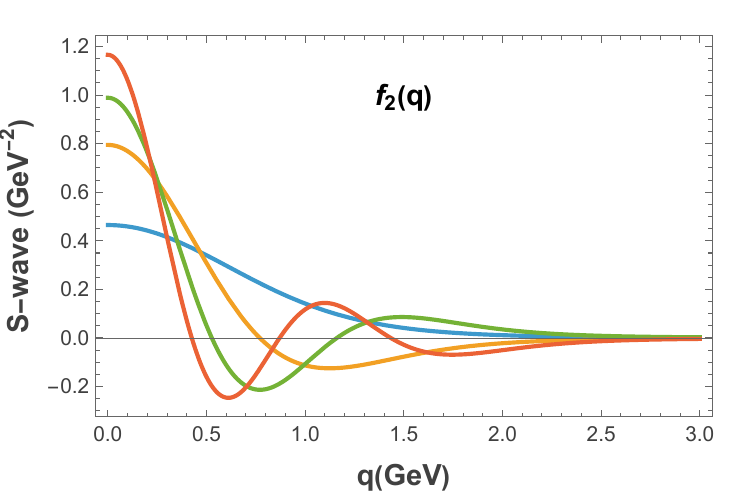}
\caption{Radial components $f_1(q)$ and $f_2(q)$ of the Salpeter wave function for the S-wave $J^P=0^-$ $t\bar{c}$ system. The blue, orange, green, and red curves correspond to the first four radial excitations.} \label{figtc0-}
\end{figure}

\begin{figure}[H]
\includegraphics[width=6.8cm]{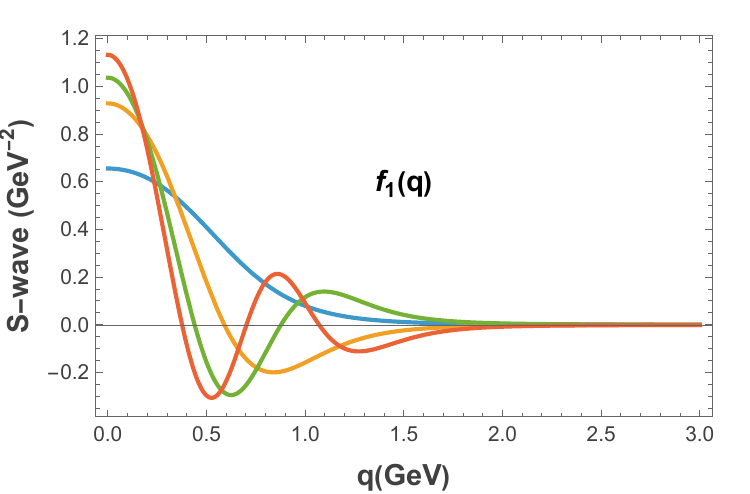}
\includegraphics[width=6.8cm]{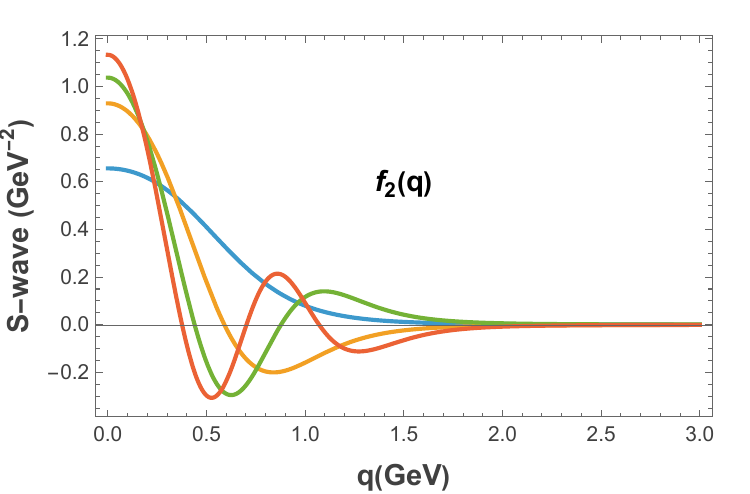}
\caption{Radial components $f_1(q)$ and $f_2(q)$ of the Salpeter wave function for the $J^P=0^-$ S-wave $t\bar{s}$ system. The blue, orange, green, and red curves correspond to the first, second, third, and fourth radial excitations, respectively.} \label{figts0-}
\end{figure}

\begin{figure}[H]
\includegraphics[width=6.8cm]{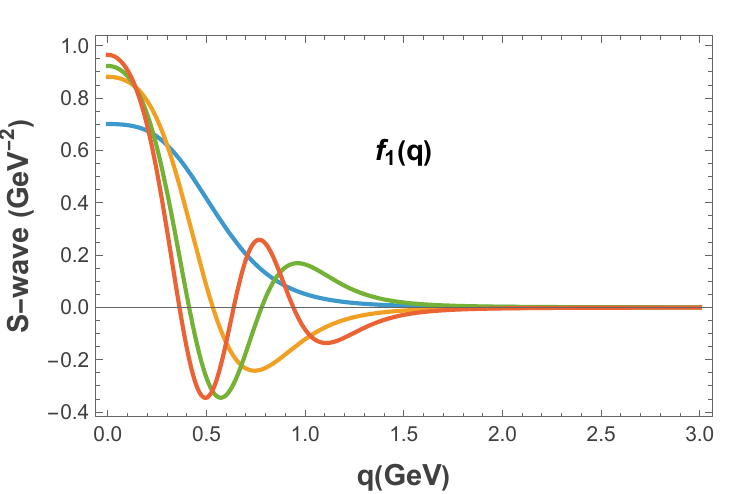}
\includegraphics[width=6.8cm]{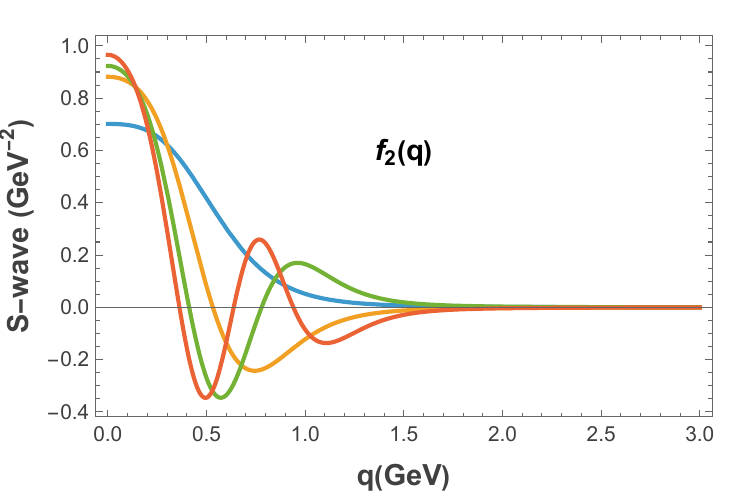}
\caption{The radial components $f_1(q)$ and $f_2(q)$ of the Salpeter wave function for the S-wave $J^P=0^-$ $t\bar{d}$ system are shown. The blue, orange, green, and red curves correspond to the first four radial excitations.} \label{figtd0-}
\end{figure}

\begin{figure}[H]
\includegraphics[width=6.8cm]{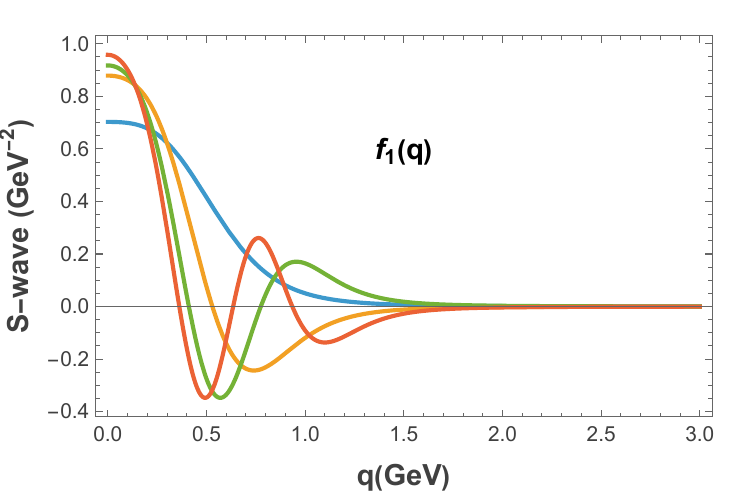}
\includegraphics[width=6.8cm]{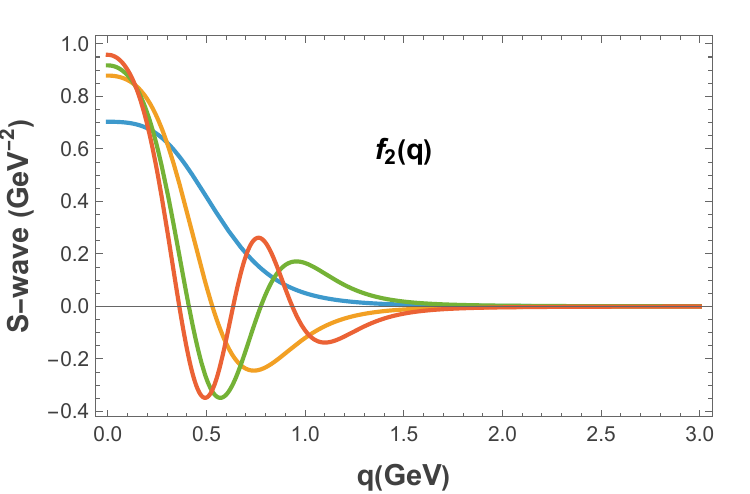}
\caption{Radial components $f_1(q)$ and $f_2(q)$ of the Salpeter wave function for the S-wave $J^P=0^-$ $t\bar{u}$ system. The blue, orange, green, and red curves correspond to the first four radial excitations, respectively.} \label{figtu0-}
\end{figure}

\end{widetext}

\end{document}